\documentclass[a4paper,11pt]{article}
\usepackage{pos}
\usepackage{xspace}
\usepackage{bbm}

\newcommand{\repoURL}{https://github.com/evanberkowitz/supervillain}

\newcommand{\Figref}[1]{Figure~\ref{fig:#1}\xspace}
\renewcommand{\eqref}[1]{(\ref{eq:#1})\xspace}

\newcommand{\eq}[1]{#1 \eqref{#1}}
\newcommand{\doi}[1]{\href{http://doi.org/#1}{[#1]}}

\newcommand{\Integers}{\mathbb{Z}\xspace}

\newcommand{\Reals}{\mathbb{R}\xspace}

\usepackage{multirow}

\newcommand{\supervillain}{\texttt{supervillain}\xspace}

\let\builtinLaTeX\LaTeX
\def\LaTeX{\builtinLaTeX\xspace}

\title{A Vector-Vector-Axial Anomaly in 4D}
\ShortTitle{A VVA Anomaly in 4D}

\author*[a]{Evan Berkowitz}
\author[b,c]{Shi Chen}
\author[b]{Aleksey Cherman}

\affiliation[a]{University of the Virgin Islands,\\
10000 Castle Burke, Kingshill VI 00850, USVI}

\affiliation[b]{School of Physics and Astronomy, University of Minnesota\\
Minneapolis, MN 55455, USA}

\affiliation[c]{Department of Physics, University of California San Diego\\
9500 Gilman Drive, La Jolla, CA 92093-0319, USA}

\emailAdd{evan.berkowitz@uvi.edu}
\emailAdd{s.chern.phys@gmail.com}
\emailAdd{acherman@umn.edu}

\abstract{Because it offers exactly-integer topological quantities, the modified Villain discretization offers a route to correct anomaly structure at finite lattice spacing. I will present a 4-dimensional purely-bosonic modified Villain model with an ABJ-type vector-vector-axial anomaly.

The model may have a direct continuous phase transition from one SSB phase to another, giving hope of yielding 4D CFT with a VVA anomaly.
}

\FullConference{The 43rd International Symposium on Lattice Field Theory (Lattice 2026)\\
July 26 to August 1, 2026\\
University of Maryland, College Park, USA\\}

\begin{document}
\maketitle

Massless QED has a \emph{chiral}, \emph{ABJ}, or \emph{vector-vector-axial} (VVA) anomaly
\begin{align}
	\partial_\mu j^\mu_A &= - \frac{1}{16\pi^2} \epsilon^{\mu\nu\rho\sigma} F_{\mu\nu} F_{\rho\sigma}
	&\text{or}&
	&
	d \star j_A &= -\frac{1}{4\pi^2} F \wedge F
	\label{eq:anomaly}
\end{align}
leveraging the notation of exterior calculus.
The same feature appears in QCD, which explains the much faster decay of the neutral pion compared to the charged pions.
The anomaly arises from a Feynman diagram with a fermion loop with two vector and one axial operator insertion.
Despite this perturbative description, the \eq{anomaly} is an exact, nonperturbative property of the Standard Model.

Preserving the chiral \eq{anomaly} in any nonperturbative definition of the Standard Model, including lattice definitions, presents a number of well-known technical challenges rooted in the Nielsen--Ninomiya theorem.
A discretization satisfying the Ginsparg--Wilson relation carries an exact lattice chiral symmetry and an exact index theorem; otherwise the anomaly is recovered only in the continuum limit.
Neither route yet yields a nonperturbative construction of a nonabelian chiral gauge theory, though a modified Villain construction gives an abelian chiral gauge theory in two dimensions~\cite{Berkowitz:2023pnz}.

In this work we will present and investigate a modified Villain model~\cite{Sulejmanpasic:2019ytl,Gorantla:2021svj} which enjoys a vector-vector-axial \emph{mixed 't Hooft anomaly} exactly even at finite lattice spacing.
Interestingly, the model contains exclusively bosonic degrees of freedom.
The model may contain a continuous phase transition, yielding a 4D CFT with the chiral anomaly, but preliminary numerical studies have not yet yielded a clear picture.

\section{Background: The Compact Boson + XY Model}

The usual discretization of the XY model
\begin{align}
	S_{\text{Wilsonian}} = \kappa \sum_{\ell} \left[1- \cos (d\varphi)_\ell \right]
	\label{eq:Wilsonian construction}
\end{align}
contains a bosonic field $\varphi$ defined on the sites of a lattice, with $d$ the simplest nearest-neighbor `forward' derivative that lives on the lattice links $\ell$.
In four dimensions we focus on the model formulated on the standard hypercubic lattice.
The model has a global shift symmetry and a local $2\pi$ periodicity on every site $s$,
\begin{align}
	\text{global shift}: \quad \varphi&\to\varphi+c
	&
	\text{local shift}: \quad \varphi_s&\to\varphi_s + 2\pi.
\end{align}

In contrast, the \emph{Villain} discretization~\cite{Villain:1974ir}
\begin{align}
	S_{\text{Villain}} &= \frac{\kappa}{2} \sum_\ell \left[ (d\varphi)_\ell - 2\pi n_\ell \right]^2
	\label{eq:Villain model}
\end{align}
uses a real-valued $\varphi$ and an integer-valued link field $n$ and replaces the local shift symmetry with the discrete gauge redundancy
\begin{align}
	\varphi_s &\to \varphi_s + 2\pi k_s
	&
	n_\ell &\to n_\ell + (dk)_\ell
	&
	(k_s &\in \Integers)
\end{align}
which replicates the $U(1)$ symmetry of the \eq{Wilsonian construction} since $U(1) \cong \Reals / 2\pi \Integers$.

The \eq{Villain model} has a number of nice features.
For example, because it is quadratic one can exhibit exact lattice dualities by Poisson resummation~\cite{Sulejmanpasic:2019ytl}.
Topological quantities also come out to be integers without any fancy footwork, so long as one defines the derivative of a link variable to be the obvious oriented sum that satisfies $d(d\varphi) = 0$.
For example, the lattice analog of the winding number $w_A = \frac{1}{2\pi} \oint_{\partial A} \omega = \frac{1}{2\pi} \int_A d\omega$ around any area $A$ on the lattice is
\begin{align}
	w_A = \frac{1}{2\pi}\sum_{\ell \in \partial A} (d\varphi-2\pi n)_\ell = \frac{1}{2\pi} \sum_{p\in A} \left[d(d\varphi - 2\pi n)\right]_p = - \sum_{p\in A} (dn)_p \in \Integers.
\end{align}

In fact, we may implement, in any dimension $D$, the exterior derivative $d$, coderivative $\delta$, Laplacian $\Delta$, Hodge star $\star$, and wedge $\wedge$ that satisfy the continuum relations\footnote{For experts, the lattice wedge product $\wedge$ is really the cup product $\cup$~\cite{Jacobson:2023cmr}.}
\begin{align}
	d^2 &= 0
		&
			\sum a \wedge \star b &= \left\langle a, b \right\rangle
	\nonumber\\
	\left\langle da, b\right\rangle &= \left\langle a, \delta b\right\rangle
		&
			(a+b) \wedge c &= a \wedge c + b \wedge c
	\nonumber\\
	\delta^2 &= 0
		&
			(a \wedge b) \wedge c &= a \wedge (b \wedge c)
	\nonumber\\
	\Delta &= d\delta + \delta d
		&
			d(a \wedge b) &= da \wedge b + (-1)^{p} a \wedge db
	\nonumber\\
	\left\langle \Delta f, f \right\rangle &= \left\langle df, df \right\rangle + \left\langle \delta f, \delta f \right\rangle
\end{align}
for any $p$-form $a$ and $q$-form $b$ and the $\langle \cdot, \cdot\rangle$ notation is the natural lattice inner product.
However, three continuum relations break, and only two can be exactly repaired,
\begin{align}
	a \wedge b &= (-1)^{pq} b \wedge a
	&\to&&
	a \wedge b &\neq (-1)^{pq} b \wedge a
	\\
	\star \star a &= (-1)^{p(D-p)} a
	&\to&&
	\star \star a &= (-1)^{p(D-p)} {\color{red} T} a
	\\
	\delta &= (-1)^{D(k+1)+1} \star d \star
	&\to&&
	\delta &= (-1)^{D(k+1)+1} {\color{red} T} \star d \star
\end{align}
where $T$ is a fixed lattice translation and $k$ is the degree of the form $\delta$ acts on.
On the hypercubic lattice the inability to maintain all three of these continuum relations is a lattice exterior calculus statement analogous to the Nielsen--Ninomiya theorem.

As written, the \eq{Villain model} in $D=4$ dimensions exhibits a quantum phase transition as a function of the single dimensionless parameter $\kappa$.
A good probe of this transition is the spin two-point correlation function
\begin{align}
	C_\varphi(\Delta x) &= \left\langle e^{+i\varphi_{x+\Delta x}} e^{-i\varphi_x}\right\rangle.
	\label{eq:spin two-point correlation function}
\end{align}
At large coupling $\kappa$ the model spontaneously breaks the $U(1)_\varphi$ symmetry and exhibits spin order so that the two-point correlation function does not vanish in the long-distance limit.
In contrast, at low coupling the vortices proliferate, $U(1)_\varphi$ is restored, and the two-point function is gapped,
\begin{align}
	\lim_{\Delta x \to \infty} C_\varphi(\Delta x) & \sim \begin{cases}
		e^{-m \Delta x} & \kappa \text{ small}
		\\
		\text{constant} & \kappa \text{ large.}
	\end{cases}
	\label{eq:unconstrained phase diagram}
\end{align}

\section{Jacobson's No-Intersections Model}

A forthcoming paper by Theo Jacobson~\cite{Jacobson:toappear} observes that on the four-dimensional hypercubic lattice the model
\begin{align}
	S_{\text{NI}} = \frac{\kappa}{2} \sum_{\ell} (d\varphi - 2\pi n)_\ell^2 + i \sum_h \theta_h (dn \wedge dn)_h
	\label{eq:no-intersections model}
\end{align}
with a real-valued Lagrange multiplier field $\theta$ on every hypercube has two $U(1)$ global symmetries,
\begin{align}
	U(1)_V: \quad \varphi &\to \varphi + \alpha
	&
	U(1)_A: \quad \theta &\to \theta + \beta
	\label{eq:global symmetries}
\end{align}
which have a mixed 't Hooft anomaly on the lattice
\begin{align}
	d\star j_A &= \frac{1}{4\pi^2} F_V \wedge F_V.
\end{align}
This model looks like it has a horrible sign problem.
But, in fact, the Lagrange multiplier term may be integrated out,
\begin{align}
	\mathcal{Z}
	= \int\!\!\!\!\!\!\!\!\sum\; D\varphi\; Dn\; D\theta\; e^{-S_\text{NI}} 
	= \int\!\!\!\!\!\!\!\!\sum\; D\varphi\; Dn\; e^{-S_{\text{Villain}}} \prod_h \left[(dn \wedge dn)_h = 0 \right]
	\label{eq:no-intersections constraint}
\end{align}
where the product's factors are Iverson brackets that set the vortex worldsheet intersection density
\begin{align}
	q = dn \wedge dn
	\label{eq:intersection density}
\end{align}
to zero everywhere.
In other words, the model's valid configurations are a subset of the configurations of the unconstrained \eq{Villain model}.
In the continuum this can be interpreted as a prohibition of the intersection of vortex worldsheets, so we call this Jacobson's no-intersection model.

The remainder of this work will be an exploration of the no-intersection model to try and understand its phase diagram.
At large $\kappa$ the unmodified \eq{Villain model} has spontaneous $U(1)_\varphi$ breaking and hardly ever enjoys $dn\neq0$ since vortices are rare.
Therefore we expect that the no-intersection constraint of Jacobson's model should hardly matter at large $\kappa$: we still expect $U(1)_\varphi$ to spontaneously break.

But what about low $\kappa$?
Is $U(1)_\varphi$ broken regardless of $\kappa$?
Is it restored, as in the \eq{unconstrained phase diagram}?
If so, we must expect $U(1)_\theta$ to be broken, to carry the anomaly.
Is there a single, direct, continuous transition in this case?
A single but first-order transition?
More than one transition?
Could there be a whole line of CFTs at small $\kappa$?

As a first check we can see if the constraint matters for the existing XY transition.
If the \eq{intersection density} is substantial at the transition of the unconstrained \eq{Villain model}, we should expect the phase diagram to be meaningfully altered.

\begin{figure}[t]
	\includegraphics[width=0.49\textwidth]{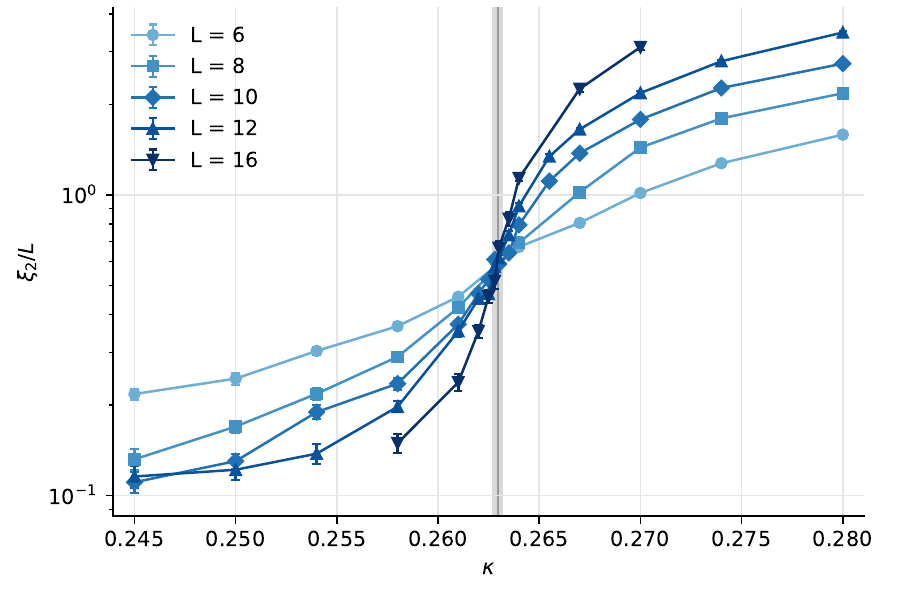}
	\includegraphics[width=0.49\textwidth]{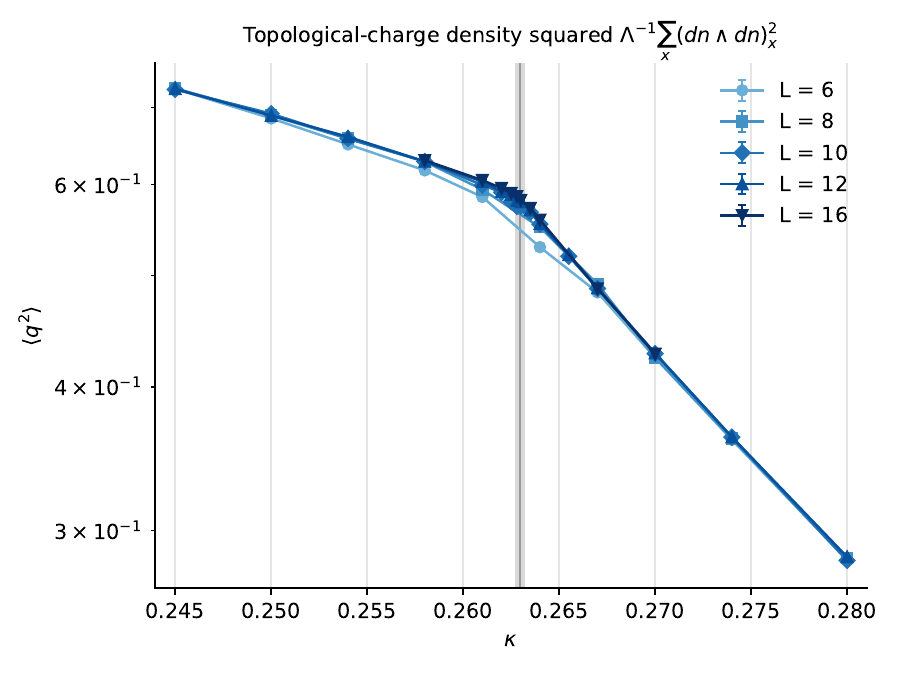}
	\caption{
		LEFT: The correlation length of the \eq{spin two-point correlation function} normalized by the lattice size $L$ as a function of $\kappa$.
		The vertical band is the estimated critical coupling of the four-dimensional XY model.
		RIGHT: The mean square intersection density $\left\langle q^2 \right\rangle$ as a function of $\kappa$.
	}
	\label{fig:Villain transition}
\end{figure}

We simulated the unconstrained four-dimensional \eq{Villain model} with \supervillain~\cite{supervillain} and measured the correlation length of the \eq{spin two-point correlation function} to determine the critical coupling $\kappa_c \approx 0.263$, as shown in the left panel of \Figref{Villain transition}.
At that same coupling we observe that the mean square intersection density $\left\langle q^2 \right\rangle$ is substantial, about $0.6$, so that the configurations important to the unconstrained model's transition very often violate the \eq{no-intersections constraint}.
Therefore, the no-intersection model should behave quite differently as a function of $\kappa$.

\section{Preliminary Results}

Sampling the valid constraint-satisfying configurations of Jacobson's \eq{no-intersections model} is a challenging exercise.
In particular, while worm algorithms can sample some modified Villain constructions~\cite{Berkowitz:2024iuv}, they struggle with Jacobson's no-intersections model because the constraint is quadratic in $n$.
On nontrivial vortex configurations it can be very difficult to find an update that moves an intersection defect without also populating nearby hypercubes with additional unintended constraint violations.

So, we can study the model as-is with the hard \eq{no-intersections constraint} or we can soften the constraint by introducing a fugacity for the defects,
\begin{align}
	\mathcal{Z}_{\text{soft}}
	= \int\!\!\!\!\!\!\!\!\sum\; D\varphi\; Dn\; e^{-S_{\text{Villain}}} \zeta^{\sum_h \left|dn \wedge dn\right|},
	\label{eq:soft-constraint model}
\end{align}
so that the model is a grand-canonical gas of intersection defects.
When the fugacity $\zeta$ is 1 the intersections are totally unpunished and the model is identical to the unconstrained \eq{Villain model}, while in the limit $\zeta\to0$ the soft model reproduces the \eq{no-intersections constraint}.
Not knowing which route will be easiest, we study both the model with the hard constraint and the softly-constrained model and try to take the zero-fugacity limit as best we can.
At $\zeta > 0$ the $U(1)_A$ \eqref{global symmetries} is explicitly broken and there is no anomaly to worry about, but it is recovered when the fugacity vanishes.

\begin{figure}[t]
	\begin{center}
	\begin{tabular}{ccc}
		\includegraphics[width=0.3\textwidth]{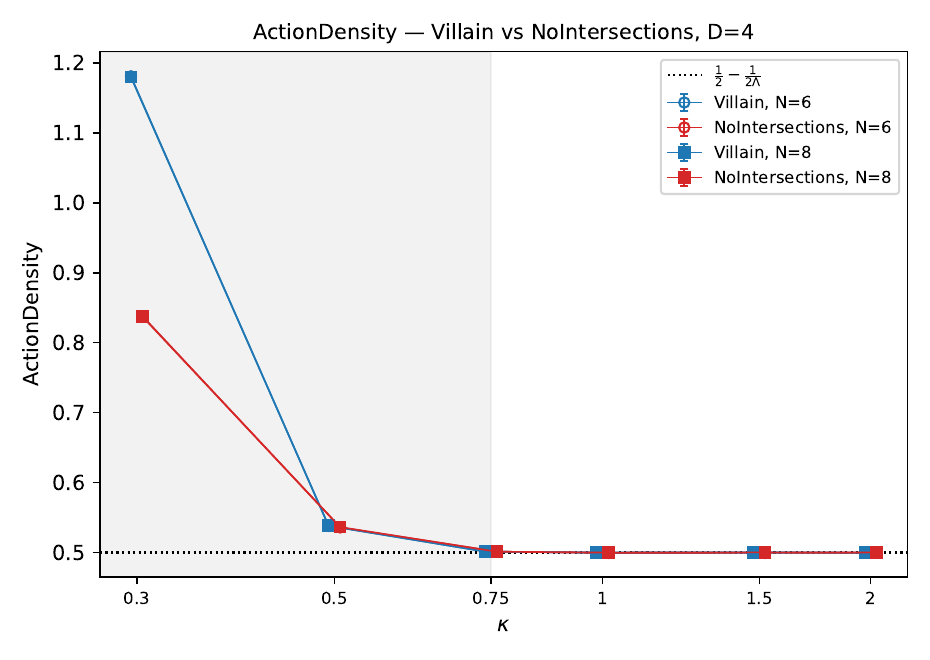}
		&
		\includegraphics[width=0.3\textwidth]{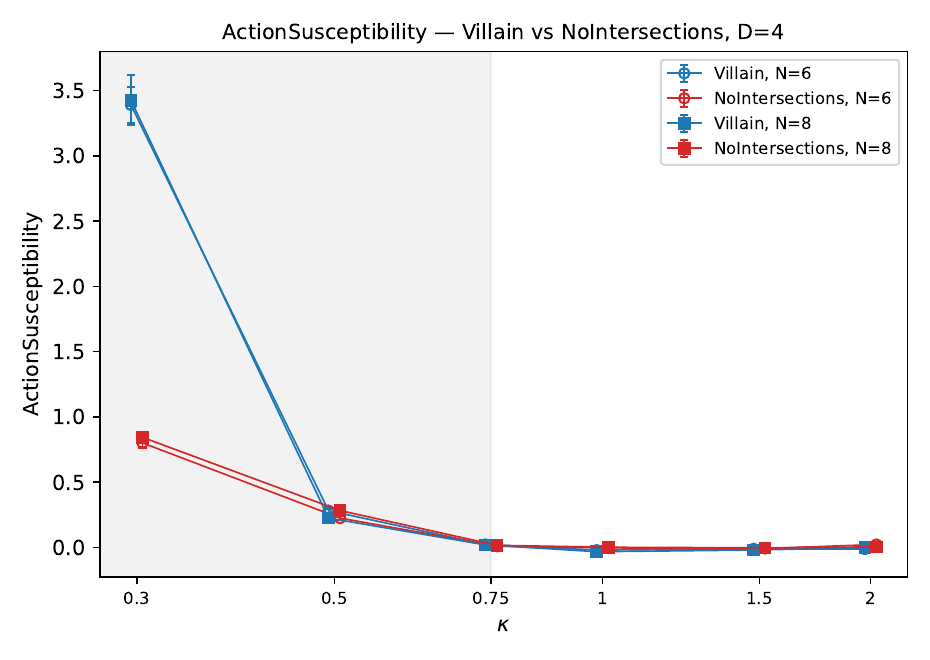}
		&
		\includegraphics[width=0.3\textwidth]{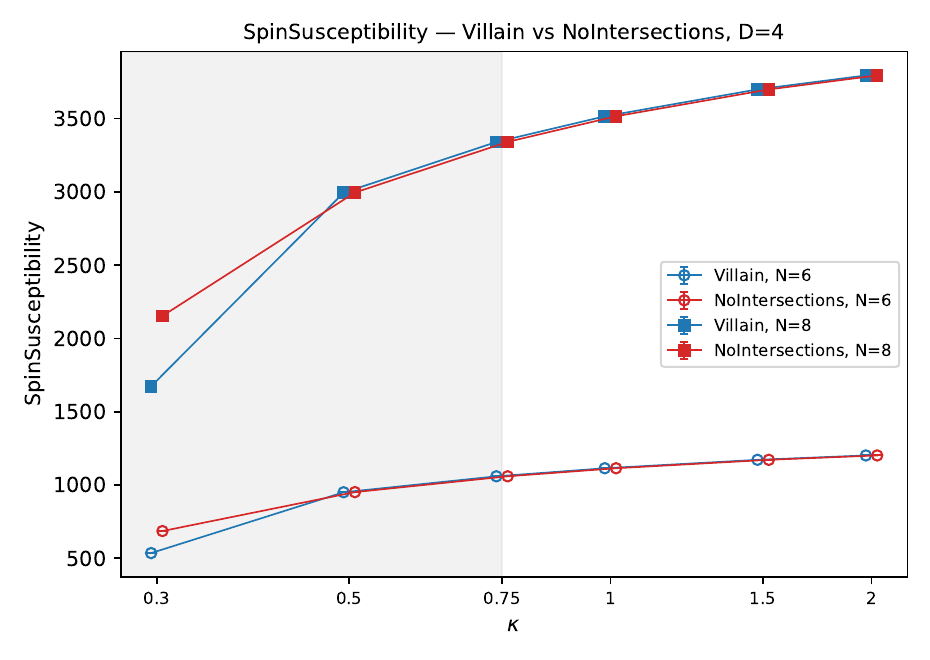}
		\\
		\includegraphics[width=0.3\textwidth]{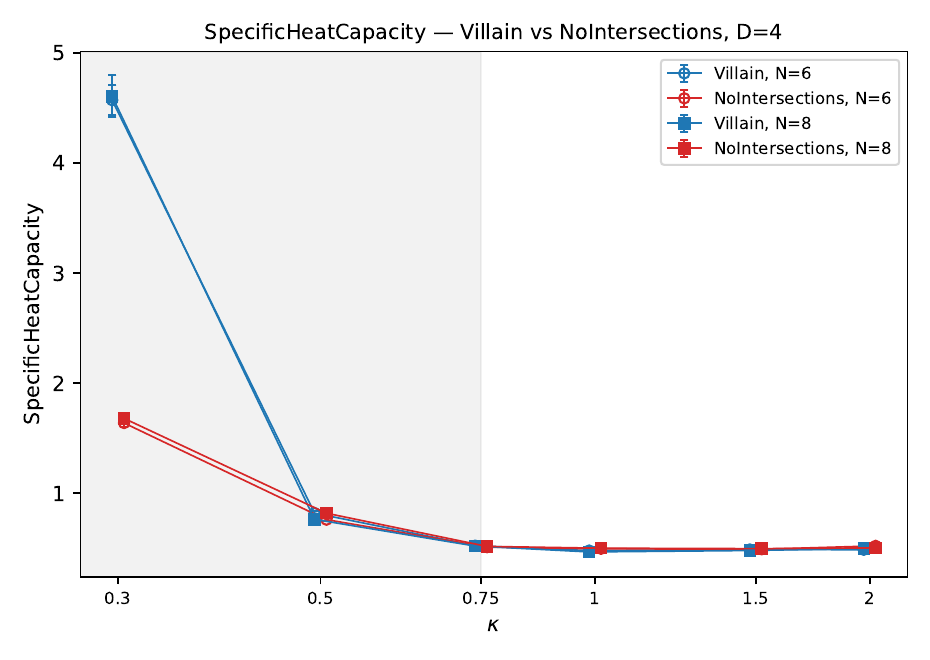}
		&
		\multicolumn{2}{c}{\includegraphics[width=0.6\textwidth]{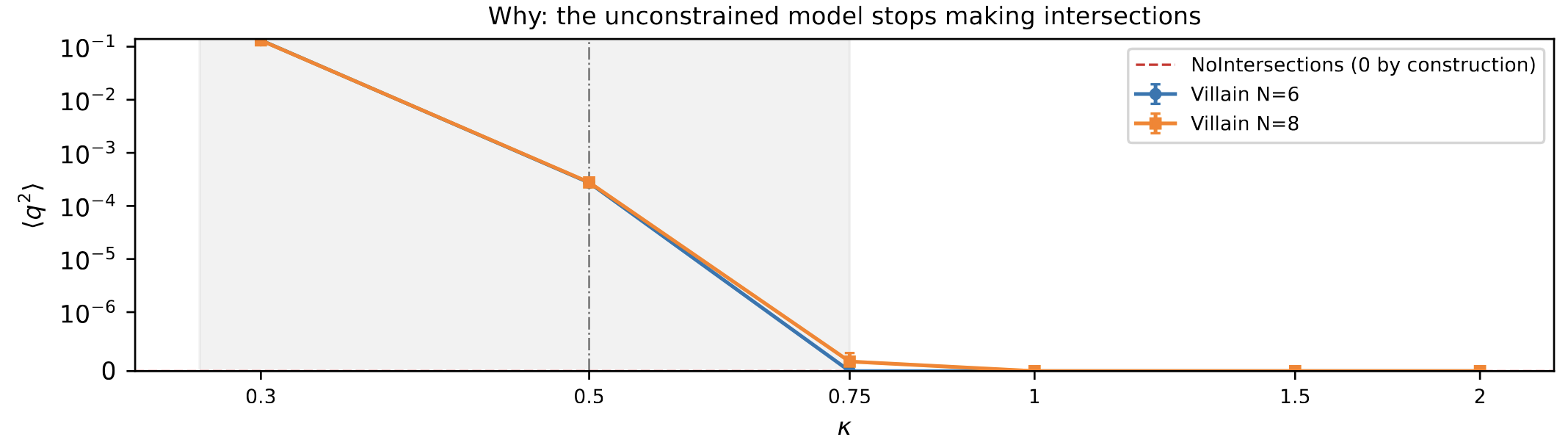}}
	\end{tabular}
	\end{center}
	\caption{
		Five observables as a function of $\kappa$.
		In the first four panels the unconstrained \eq{Villain model} results are shown in blue and the \eq{no-intersections model} results are shown in red, with open markers indicating a $6^4$ and filled markers an $8^4$ lattice.
		Across the first row and into the second we have the action density, action susceptibility, spin susceptibility, and specific heat capacity, which all agree at large coupling $\kappa$ and meaningfully diverge below $\kappa \sim 0.75$ (gray band).
		The last extra-wide panel offers the explanation: that $\kappa$ is where intersection production begins.
	}
	\label{fig:matching}
\end{figure}

At large coupling $\kappa$ the unconstrained model hardly ever carries vortices or their worldsheet intersections, and we expect good agreement with the constrained model.
In \Figref{matching} we study the $\kappa$ dependence of five observables in the unconstrained \eq{Villain model} (blue) and with the \eq{no-intersections constraint} (red).
The observables all match at $\kappa \gtrsim 0.75$ but diverge at lower couplings.
The double-wide panel offers the explanation: that coupling is where the unconstrained model first produces meaningful intersection density, and below that coupling the configurations the two models value meaningfully differ.

Perhaps the most useful handles for understanding the phase structure are the \eq{spin two-point correlation function}, the analogous two-point function of the Lagrange multiplier $\theta$
\begin{align}
	C_\theta(\Delta x) = \left\langle e^{+i\theta_{x+\Delta x}} e^{-i\theta_x} \right\rangle
	\label{eq:defect two-point correlation function}
\end{align}
and the corresponding susceptibilities
\begin{align}
	\chi = \sum_{\Delta x} C(\Delta x).
\end{align}
While the Lagrange multiplier $\theta$ is not present in the simulations---it is integrated out in order to eliminate the sign problem---we can measure its two-point function with update algorithms that allow defects to propagate, as is often done in models simulated with worm algorithms.

\begin{figure}
	\raisebox{-0.5\height}{\includegraphics[width=0.45\textwidth]{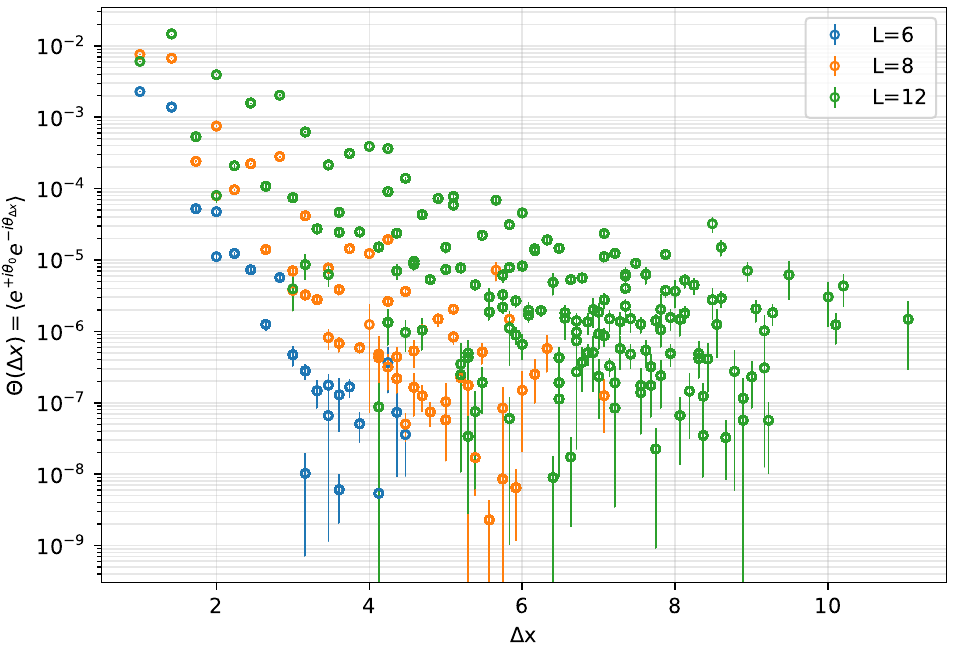}}
	\raisebox{-0.5\height}{\includegraphics[width=0.45\textwidth]{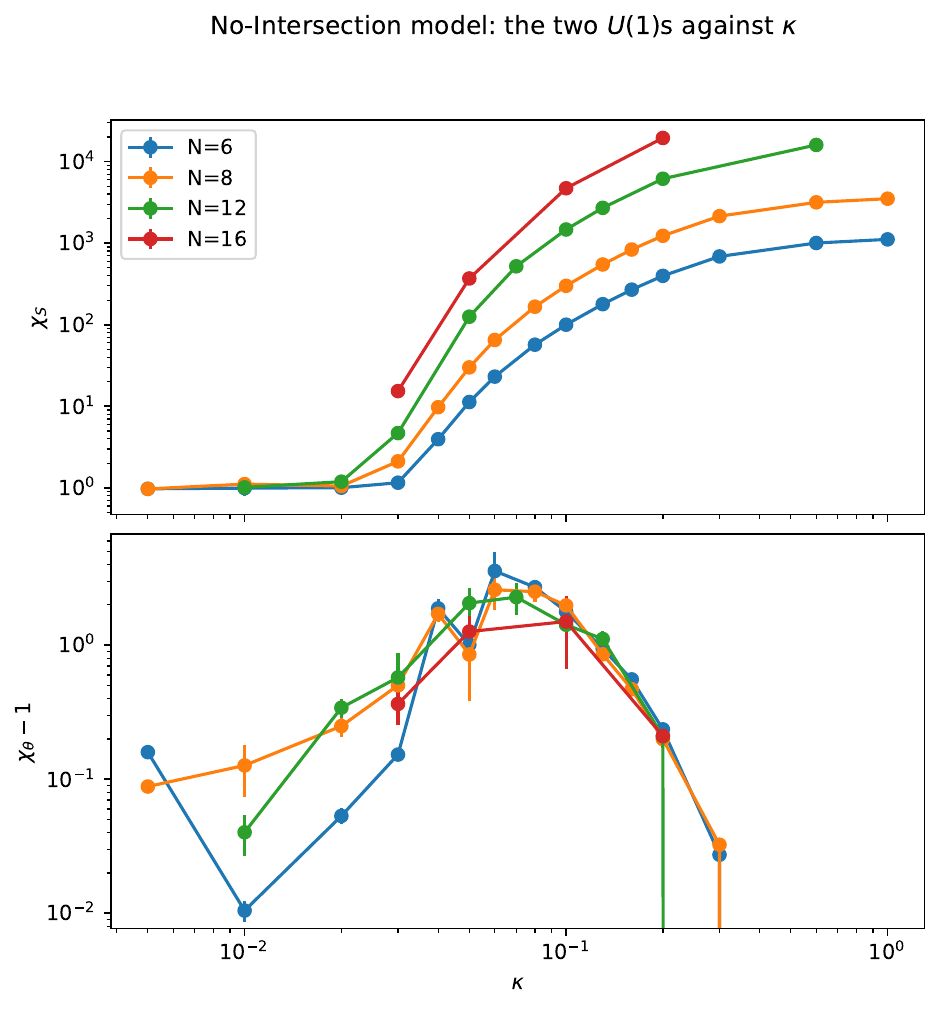}}
	\caption{
		LEFT: The $\theta$ two-point function in the no-intersections model for three different lattices at $\kappa=0.02$.
		RIGHT: On top, the $\varphi$ susceptibility, the integral of the \eq{spin two-point correlation function}, in the no-intersections model for four different lattice sizes; below, the analogous $\theta$ susceptibility.
	}
	\label{fig:susceptibility trade}
\end{figure}

In the left panel of \Figref{susceptibility trade} we show the \eq{defect two-point correlation function} at $\kappa=0.02$ for three different lattices.
We observe the correlator's tail rises as the lattice grows, though we must concede we have not seen a clear long-distance plateau.
In the right panel we show the spin and defect susceptibilities.
At large coupling the spin susceptibility is large and shows the fanning typical in a spontaneously-broken phase, and the defect susceptibility is small.
At small coupling the spin susceptibility plummets but the defect susceptibility does not fan.
We suspect this is due to sampling difficulties rather than physics, but without a cure we cannot say for sure.

As it stands, the small-coupling behavior is ambiguous.
But one way or the other it cannot be that both susceptibilities crash---that would violate the anomaly condition.

\begin{figure}
	\includegraphics[width=0.9\textwidth]{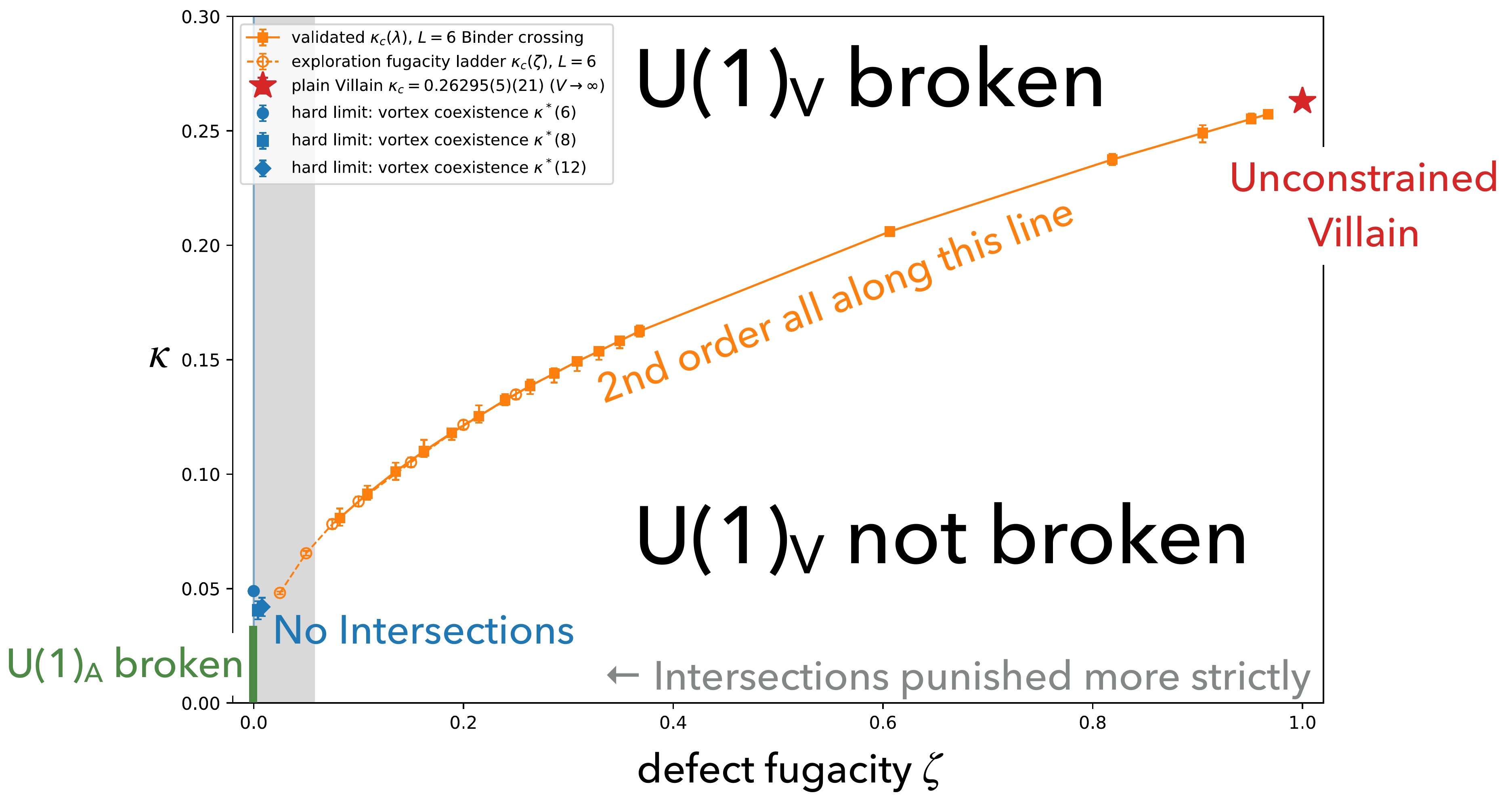}
	\caption{
		The phase diagram of the \eq{soft-constraint model}.
		A phase with spontaneous $U(1)_V$ breaking at large coupling is separated from a $U(1)_V$-symmetric phase at low coupling by a line (orange) of continuous transitions.
		As the defect fugacity $\zeta$ vanishes sampling the soft-constraint model becomes more difficult and our results are not yet reliable (gray band).
		Also shown is the estimated critical coupling of the \eq{no-intersections model} based on the spin susceptibility.
		The possible line (green) with spontaneously-broken $U(1)_A$ symmetry emerges at $\zeta=0$.
	}
	\label{fig:soft phase diagram}
\end{figure}

As mentioned, we also studied the soft-constraint model, where rather than prohibit intersections completely we price them instead.
In \Figref{soft phase diagram} we summarize our results.
At intersection fugacity $\zeta=1$ we recover the phase structure of the unconstrained \eq{Villain model} (red star).
As we reduce the fugacity we still find, by studying the usual Binder cumulant, a continuous phase transition (orange) separating a $U(1)_V$ spontaneous-breaking phase from a restored phase where $U(1)_V$ is not broken.
Perhaps unsurprisingly, as we make the fugacity smaller and smaller and the model recovers the \eq{no-intersections constraint} the sampling becomes more and more costly---a conservation of difficulty.
In exactly the most interesting part of the parameter space---small fugacity (the gray band)---the results become unreliable.
However, it is encouraging that our estimate of the critical coupling in the no-intersections model and the extrapolated phase transition line appear compatible.

\section{Summary and Outlook}

Using \supervillain we studied Theo Jacobson's \eq{no-intersections model}, a purely bosonic modified Villain XY model that exhibits a mixed vector-vector-axial 't Hooft anomaly in four dimensions, together with a model with a softer constraint \eqref{soft-constraint model} that interpolates between the no-intersections model and the unconstrained \eq{Villain model}.
The no-intersections model matches the unconstrained XY model at large coupling, but at low coupling, where vortices proliferate in the unconstrained model, it seems quite different.
Despite suspected ergodicity and sampling difficulties, it seems possible that the no-intersections model undergoes a single continuous transition.
If so, it would be exciting to understand the physics of the corresponding ABJ-anomaly-carrying conformal field theory.

\section*{Acknowledgements}

The authors thank Theodore Jacobson for explaining the details of his no-intersections model and Theodore Jacobson and Shu-Heng Shao for useful discussions.
EB was supported by the National Aeronautics and Space Administration under Award Number 80NSSC22M0063 as part of the VI NASA EPSCoR program.
AC and SC were supported by the Simons Foundation through the Collaboration on Confinement and QCD Strings under award number 994302.

\bibliographystyle{unsrtnat}
\bibliography{master}

\end{document}